\documentclass[oneside,english]{elsart}
\usepackage[T1]{fontenc}
\usepackage[latin1]{inputenc}
\usepackage{amsmath}
\usepackage{graphicx}
\usepackage{amssymb}

\makeatletter

\providecommand{\tabularnewline}{\\}

\usepackage{latexsym}
\usepackage{ae,aecompl}
\usepackage{graphicx}
\usepackage{amsfonts}

\usepackage{babel}
\makeatother
\begin{document}
\begin{frontmatter}
\title{Fast spin $\pm2$ spherical harmonics transforms and application in cosmology}
\author[yp]{Y. Wiaux\corauthref{cor}},
\corauth[cor]{Corresponding author.}
\ead{yves.wiaux@epfl.ch}
\author[l]{L. Jacques},
\author[yp]{P. Vandergheynst}
\address[yp]{Signal Processing Institute, Ecole Polytechnique Fédérale de Lausanne (EPFL), CH-1015 Lausanne, Switzerland}
\address[l]{Communications and Remote Sensing Laboratory, Université catholique de Louvain (UCL), B-1348 Louvain-la-Neuve, Belgium}
\begin{abstract}
A fast and exact algorithm is developed for the spin $\pm2$ spherical harmonics transforms on equi-angular pixelizations on the sphere. It is based on the Driscoll and Healy fast scalar spherical harmonics transform. The theoretical exactness of the transform relies on a sampling theorem. The associated asymptotic complexity is of order $\mathcal{O}(L^2 \log^{2}_{2}L)$, where $2L$ stands for the square-root of the number of sampling points on the sphere, also setting a band limit $L$ for the spin $\pm2$ functions considered. The algorithm is presented as an alternative to existing fast algorithms with an asymptotic complexity of order $\mathcal{O}(L^3)$ on other pixelizations. We also illustrate these generic developments through their application in cosmology, for the analysis of the cosmic microwave background (CMB) polarization data.
\end{abstract}
\end{frontmatter}

\section{Introduction}

In the last few years, the analysis of the temperature anisotropies
of the cosmic microwave background (CMB), together with other cosmological
observations, has allowed the definition of a precise concordance
cosmological model. These observations culminated with the release
of the three-year data of the Wilkinson Microwave Anisotropy Probe
(WMAP) satellite experiment. The cosmological parameters are now determined
with an unprecedented precision of the order of several percent \cite{CMBbennett,CMBhinshaw,CMBspergel1,CMBspergel2}.
In the concordance model, the CMB originates from quantum energy fluctuations
defined in a primordial era of inflation. These tiny fluctuations
are Gaussian in first approximation. The cosmological principle of
homogeneity and isotropy of the universe is also assumed. The observed
radiation is therefore understood as a unique realization of a Gaussian
and stationary (\emph{i.e.} homogeneous and isotropic) random process
on the sphere, which may be completely characterized from its two-point
correlation functions, or the corresponding angular power spectra. 

The present concordance values of the cosmological parameters are
obtained through a best fit of the theoretical temperature angular
power spectrum of the CMB with the experimental data. Beyond temperature
anisotropies, \emph{i.e.} intensity anisotropies, a polarization of
the CMB is also present which constitutes a complementary source of
information for cosmology. This polarization is produced through Thomson
scattering at the epoch of recombination. The degree of polarization
of the CMB is expected to be of the order of $10$ percent of the
temperature anisotropies at small scales, and lower at large scales.
As Thomson scattering only produces linearly polarized light, the
CMB radiation is completely described by its temperature $T$, and
its linear polarization Stokes parameters $Q$ and $U$ \cite{CMBPOLkosowsky1,CMBPOLzaldarriaga,CMBPOLseljak,CMBPOLkamionkowski1,CMBPOLkamionkowski2,CMBPOLhu1}.
First polarization measurements were recently obtained, notably by
the WMAP experiment \cite{CMBpage}. Future CMB experiments such as
the Planck surveyor satellite experiment will allow a deeper probe
of the temperature and polarization spectra, thanks to improved sensitivity
and resolution on the whole sky.

From the mathematical point of view, the observable temperature $T$
is a scalar function on the sphere, \emph{i.e.} invariant under local
rotations in the plane tangent to the sphere at each point. The associated
invariant $TT$ angular power spectrum results from the decomposition
of the temperature in scalar spherical harmonics. But the observable
polarization Stokes parameters $Q$ and $U$ transform as the components
of a transverse, symmetric, and traceless rank $2$ tensor under local
rotations. However, scalar electric $E$ and magnetic $B$ polarization
components may equivalently be defined from the parameters $Q$ and
$U$. The associated invariant $EE$ and $BB$ polarization angular
power spectra, and the cross-correlation $TE$ spectrum result from
the decomposition of the combinations $Q\pm iU$ in spin $\pm2$ spherical
harmonics on the sphere \cite{CMBPOLzaldarriaga}. From the numerical
point of view, the asymptotic complexity associated with a naive quadrature
based on the definition of the scalar and spin $\pm2$ spherical harmonics
transforms is of order $\mathcal{O}(L^{4})$, where $L$ roughly identifies
the square-root of the number of sampling points on the sphere. Corresponding
computation times for the analysis of megapixels all-sky maps such
as those of the ongoing WMAP or the forthcoming Planck experiments
are of the order of days. Fast and precise computation methods for
the scalar and spin $\pm2$ spherical harmonics transforms of functions
on the sphere are therefore needed.

Beyond cosmology, an algorithm for the spin $\pm2$ spherical harmonics
transforms will find application in the spectral analysis of arbitrary
spin $\pm2$ signals on the sphere, components of transverse, symmetric,
and traceless rank $2$ tensor fields under local rotations.

In the present work we develop a fast algorithm for the spin $\pm2$
spherical harmonics transforms of band-limited functions on the sphere.
It is based on an existing fast algorithm for the scalar spherical
harmonics transform. It is defined on $2L\times2L$ equi-angular pixelizations
in spherical coordinates $(\theta,\varphi)$ on the sphere. The algorithm
is theoretically exact thanks to the existence of a sampling theorem.
The associated asymptotic complexity is of order $\mathcal{O}(L^{2}\log_{2}^{2}L)$.
Corresponding computation times for megapixels maps are reduced to
seconds. The algorithm is presented as an alternative to existing
fast algorithms with an asymptotic complexity of order $\mathcal{O}(L^{3})$
on other pixelizations which are widely used in the context of astrophysics
and cosmology.

In § \ref{sec:Spin-n-functions}, we recall the notion of spin $n$
functions on the sphere. In § \ref{sec:OL2log2L-algorithm}, we define
and implement a fast and exact algorithm with complexity $\mathcal{O}(L^{2}\log_{2}^{2}L)$
for the spin $\pm2$ spherical harmonics transforms on equi-angular
pixelizations. In § \ref{sec:CMB-polarization-spectra}, we illustrate
the interest of our algorithm in the context of the analysis of CMB
polarization data. We finally briefly conclude in § \ref{sec:Conclusion}.

\section{Spin $n$ functions on the sphere}

\label{sec:Spin-n-functions}In this section, we discuss standard
harmonic analysis on the sphere and on the rotation group $SO(3)$.
We also discuss the notion of spin $n$ functions on the sphere and
their decomposition in a basis of spin-weighted spherical harmonics
of spin $n$.

\subsection{Standard harmonic analysis}

Let the function $G(\omega)$ be a square-integrable function in $L^{2}(S^{2},d\Omega)$
on the unit sphere $S^{2}$. The spherical coordinates of a point
on the unit sphere, defined in the right-handed Cartesian coordinate
system $(o,o\hat{x},o\hat{y},o\hat{z})$ centered on the sphere, read
as $\omega=(\theta,\varphi)$. The angle $\theta\in[0,\pi]$ is the
polar angle, or co-latitude. The angle $\varphi\in[0,2\pi[$ is the
azimuthal angle, or longitude. The invariant measure on the sphere
reads $d\Omega=d\cos\theta d\varphi$. The standard scalar spherical
harmonics $Y_{lm}(\omega)$, with $l\in\mathbb{N}$, $m\in\mathbb{Z}$,
and $|m|\leq l$, form an orthonormal basis for the decomposition
of functions in $L^{2}(S^{2},d\Omega)$ on the sphere \cite{SASvarshalovich}.
They are explicitly given in a factorized form in terms of the associated
Legendre polynomials $P_{l}^{m}(\cos\theta)$ and the complex exponentials
$e^{im\varphi}$ as \begin{equation}
Y_{lm}(\theta,\varphi)=\left[\frac{2l+1}{4\pi}\frac{\left(l-m\right)!}{\left(l+m\right)!}\right]^{1/2}P_{l}^{m}\left(\cos\theta\right)e^{im\varphi}.\label{1}\end{equation}
This corresponds to the choice of Condon-Shortley phase $(-1)^{m}$
for the spherical harmonics, ensuring the relation $(-1)^{m}Y_{lm}^{*}(\omega)=Y_{l(-m)}(\omega)$.
This phase is here included in the definition of the associated Legendre
polynomials \cite{SASabramowitz,SASvarshalovich}. Another convention
\cite{CVbrink} explicitly transfers it to the spherical harmonics.
Any function $G(\omega)$ on the sphere is thus uniquely given as
a linear combination of scalar spherical harmonics: $G(\omega)=\sum_{l\in\mathbb{N}}\sum_{|m|\leq l}\widehat{G}_{lm}Y_{lm}(\omega)$
(inverse transform), for the scalar spherical harmonics coefficients
$\widehat{G}_{lm}=\int_{S^{2}}d\Omega\, Y_{lm}^{*}(\omega)G(\omega)$
(direct transform), with $\vert m\vert\leq l$.

Let now $G(\rho)$ be a square-integrable function in $L^{2}(SO(3),d\rho)$
on the group $SO(3)$ of three-dimensional rotations. Any rotation
$\rho\in SO(3)$ may be explicitly given in the Euler angles parametrization
as $\rho=(\varphi,\theta,\chi)$, describing successive rotations
by the Euler angles $\chi\in[0,2\pi[$, $\theta\in[0,\pi]$, and $\varphi\in[0,2\pi[$,
around the axes of coordinate $o\hat{z}$, $o\hat{y}$, and $o\hat{z}$
respectively. The invariant measure on the rotation group reads $d\rho=d\varphi d\cos\theta d\chi$.
The Wigner $D$-functions $D_{mn}^{l}(\rho)$, with $l\in\mathbb{N}$,
$m,n\in\mathbb{Z}$, and $|m|,|n|\leq l$, are the matrix elements
of the irreducible unitary representations of weight $l$ of the rotation
group $SO(3)$, in $L^{2}(SO(3),d\rho)$. By the Peter-Weyl theorem
on compact groups, the matrix elements $D_{mn}^{l*}$ also form an
orthogonal basis in $L^{2}(SO(3),d\rho)$ \cite{SASvarshalovich}.
They are explicitly given in a factorized form in terms of the real
Wigner $d$-functions $d_{mn}^{l}(\theta)$ and the complex exponentials
$e^{-im\varphi}$ and $e^{-in\chi}$ as\begin{equation}
D_{mn}^{l}\left(\varphi,\theta,\chi\right)=e^{-im\varphi}d_{mn}^{l}\left(\theta\right)e^{-in\chi}.\label{2}\end{equation}
Any function $G(\rho)$ in $L^{2}(SO(3),d\rho)$ is thus uniquely
given as a linear combination of Wigner $D$-functions : $G(\rho)=\sum_{l\in\mathbb{N}}(2l+1)/8\pi^{2}\sum_{|m|,|n|\leq l}\widehat{G}_{mn}^{l}D_{mn}^{l*}(\rho)$
(inverse transform), with $\vert m\vert,\vert n\vert\leq l$ and where
$\widehat{G}_{mn}^{l}=\int_{SO(3)}d\rho\, D_{mn}^{l}(\rho)G(\rho)$
(direct transform) stands for the with Wigner $D$-functions coefficients.

\subsection{Spin $n$ functions}

Let us define a spin $n$ square-integrable function $_{n}G(\omega)$
in $L^{2}(S^{2},d\Omega)$ on the sphere. The Euler angles $(\varphi,\theta,\chi)$
associated with a general rotation $\rho$ in three dimensions may
also be interpreted in the reverse order as successive rotations by
$\varphi$ around $o\hat{z}$, $\theta$ around $o\hat{y}'$, and
$\chi$ around $o\hat{z}''$, where the axes $o\hat{y}'\equiv o\hat{y}'(\varphi)$
and $o\hat{z}''\equiv o\hat{z}''(\varphi,\theta)$ are respectively
obtained by the first and second rotations of the coordinate system
by $\varphi$ and $\theta$ \cite{CVbrink}. The local rotations of
the basis vectors in the plane tangent to the sphere at $\omega=(\theta,\varphi)$
are rotations around $o\hat{z}''$, therefore associated with the
third Euler angle $\chi$. Spin $n$ functions on the sphere $_{n}G(\omega)$,
with $n\in\mathbb{Z}$, are defined relatively to their behavior under
the corresponding right-handed rotations by $\chi_{0}$ as \cite{SASnewman,SASgoldberg,SAScarmeli}:\begin{equation}
_{n}G'\left(\omega\right)=e^{-in\chi_{0}}\,\,{}_{n}G\left(\omega\right).\label{3}\end{equation}
 The standard square-integrable functions on the sphere considered
above are spin $0$ or scalar functions. Let us emphasize that the
rotations considered are local transformations on the sphere around
the axis $o\hat{z}''\equiv o\hat{z}''(\varphi,\theta)$, affecting
the coordinate $\chi$ in the tangent plane independently at each
point $\omega=(\theta,\varphi)$, and according to $\chi'=\chi-\chi_{0}$.
They are to be clearly distinguished from the global rotations by
$\chi$ around $o\hat{z}$ associated with the alternative Euler angles
interpretation, which affect the coordinates of the points $\omega=(\theta,\varphi)$
on the sphere. Our sign convention in the exponential is coherent
with the definition (\ref{4}) below for the spin-weighted spherical
harmonics of spin $n$. It is opposite to the original definition
\cite{SASnewman}, while equivalent to recent notations used in the
context of the CMB analysis \cite{CMBPOLzaldarriaga,CMBPOLhu1}.

Recalling the factorized form (\ref{2}), spin functions are equivalently
defined as the evaluation at $\chi=0$ of any function in $L^{2}(SO(3),d\rho)$
resulting from an expansion for fixed index $n$ in the Wigner $D$-functions
$D_{mn}^{l}(\varphi,\theta,\chi)$. The functions $D_{mn}^{l}(\varphi,\theta,0)$
or $D_{m(-n)}^{l*}(\varphi,\theta,0)$ thus naturally define for each
$n$ an orthogonal basis for the expansion of spin $n$ functions
in $L^{2}(S^{2},d\Omega)$ on the sphere. After normalization in $L^{2}(S^{2},d\Omega)$,
the spin-weighted spherical harmonics of spin $n$ are given in a
factorized form in terms of the real Wigner $d$-functions $d_{mn}^{l}(\theta)$
and the complex exponentials $e^{im\varphi}$ as \begin{equation}
_{n}Y_{lm}\left(\theta,\varphi\right)=\left(-1\right)^{n}\sqrt{\frac{2l+1}{4\pi}}d_{m(-n)}^{l}\left(\theta\right)e^{im\varphi},\label{4}\end{equation}
with $l\in\mathbb{N}$, $l\geq\vert n\vert$, and $m\in\mathbb{Z}$,
$|m|\leq l$. In particular, the symmetry properties of the Wigner
$d$-functions \cite{SASvarshalovich} imply the generalized symmetry
relation $(-1)^{n+m}\,\,{}_{n}Y_{lm}^{*}(\omega)=\,\,{}_{-n}Y_{l(-m)}(\omega)$.
The spin $0$ spherical harmonics explicitly identify with the standard
scalar spherical harmonics for the decomposition of scalar functions:
$_{0}Y_{lm}(\omega)=Y_{lm}(\omega)$, through the relation $d_{m0}^{l}(\theta)=[(l-m)!/(l+m)!]^{1/2}P_{l}^{m}(\cos\theta)$.
Any spin $n$ function $_{n}G(\omega)$ on the sphere is thus uniquely
given as a linear combination of spin $n$ spherical harmonics : $_{n}G(\omega)=\sum_{l\in\mathbb{N}}\sum_{|m|\leq l}\,\,{}_{n}\widehat{G}_{lm}\,\,{}_{n}Y_{lm}(\omega)$
(inverse transform), for the spin-weighted spherical harmonics coefficients
$_{n}\widehat{G}_{lm}=\int_{S^{2}}d\Omega\,\,{}_{n}Y_{lm}^{*}(\omega)G(\omega)$
(direct transform), with $l\geq\vert n\vert$, and $\vert m\vert\leq l$.

Finally, spin $n\pm1$ functions may be defined from spin $n$ functions
through the action of the so-called spin raising and lowering operators
\cite{SASnewman,SASgoldberg}. The action of the spin raising $\eth$
and lowering $\bar{\eth}$ operators on a spin $n$ function $_{n}G$,
giving spin $n+1$ and $n-1$ functions respectively, is defined as

\begin{equation}
\left[\eth\,\,{}_{n}G\right]\left(\theta,\varphi\right)=\left[-\sin^{n}\theta\left(\frac{\partial}{\partial\theta}+\frac{i}{\sin\theta}\frac{\partial}{\partial\varphi}\right)\sin^{-n}\theta\,\,{}_{n}G\right]\left(\theta,\varphi\right)\label{5}\end{equation}
and\begin{equation}
\left[\bar{\eth}\,\,{}_{n}G\right]\left(\theta,\varphi\right)=\left[-\sin^{-n}\theta\left(\frac{\partial}{\partial\theta}-\frac{i}{\sin\theta}\frac{\partial}{\partial\varphi}\right)\sin^{n}\theta\,\,{}_{n}G\right]\left(\theta,\varphi\right),\label{6}\end{equation}
with, under rotation by $\chi_{0}$ in the tangent plane at $\omega=(\theta,\varphi)$:
$[\eth\,\,{}_{n}G]'(\omega)=e^{-i(n+1)\chi_{0}}[\eth\,\,{}_{n}G](\omega)$
and $[\bar{\eth}\,\,{}_{n}G]'(\omega)=e^{-i(n-1)\chi_{0}}[\bar{\eth}\,\,{}_{n}G](\omega)$.
In these terms, the spin-weighted spherical harmonics of spin $n$
are related to spin-weighted spherical harmonics of spins $n+1$ and
$n-1$ through the following relations:\begin{equation}
\left[\eth\,\,{}_{n}Y_{lm}\right]\left(\omega\right)=\left[\left(l-n\right)\left(l+n+1\right)\right]^{1/2}\,\,{}_{n+1}Y_{lm}\left(\omega\right)\label{7}\end{equation}
and\begin{equation}
\left[\bar{\eth}\,\,{}_{n}Y_{lm}\right]\left(\omega\right)=-\left[\left(l+n\right)\left(l-n+1\right)\right]^{1/2}\,\,{}_{n-1}Y_{lm}\left(\omega\right),\label{8}\end{equation}
also implying\begin{equation}
\left[\bar{\eth}\eth\,\,{}_{n}Y_{lm}\right]\left(\omega\right)=-\left(l-n\right)\left(l+n+1\right)\,\,{}_{n}Y_{lm}\left(\omega\right).\label{9}\end{equation}
The corresponding direct relation between the spin-weighted spherical
harmonics of spin $n$ and scalar spherical harmonics reads:\begin{equation}
_{n}Y_{lm}\left(\omega\right)=\left[\frac{\left(l-n\right)!}{\left(l+n\right)!}\right]^{1/2}\left[\eth^{n}\, Y_{lm}\right]\left(\omega\right),\label{10}\end{equation}
for $0\leq n\leq l$, and \begin{equation}
_{n}Y_{lm}\left(\omega\right)=\left[\frac{\left(l+n\right)!}{\left(l-n\right)!}\right]^{1/2}\left(-1\right)^{n}\left[\bar{\eth}^{-n}\, Y_{lm}\right]\left(\omega\right),\label{11}\end{equation}
for $-l\leq n\leq0$.

These relations between spin-weighted and scalar spherical harmonics
are explicitly used in § \ref{sec:OL2log2L-algorithm} for the development
of a fast direct spin $\pm2$ spherical harmonics transforms algorithm.

\section{Fast spin $\pm2$ transforms algorithm}

\label{sec:OL2log2L-algorithm}In this section, we define and implement
a fast and exact algorithm for the computation of the spin $\pm2$
spherical harmonics transforms of band-limited functions on equi-angular
pixelizations on the sphere. The algorithm is based on the relations
between spin-weighted and scalar spherical harmonics established in
the previous section.

\subsection{Pixelizations and existing $\mathcal{O}(L^{3})$ algorithms}

A $2L\times2L$ equi-angular pixelization in spherical coordinates
$(\theta,\varphi)$ is defined on points $\omega_{ij}=(\theta_{i},\varphi_{j})$
for $0\leq i,j\leq2L-1$, with a uniform discretization of the coordinates:
$\Delta\theta=\theta_{i+1}-\theta_{i}=\pi/2L$ and $\Delta\varphi=\varphi_{j+1}-\varphi_{j}=2\pi/2L$.
The specific choice $\theta_{0}=\pi/4L$, and $\varphi_{0}=0$ is
considered in the following implementations. It gives $\theta_{i}=(2i+1)\pi/4L$
and $\varphi_{j}=2j\pi/2L$, and excludes the poles of the sampling,
which can be convenient for numerical reasons. The pixels centers
are identified with the sampling points $\omega_{ij}$ defined here
above. The pixels edges are identified by meridians shifted by $\Delta\theta/2=\pi/4L$,
and parallels shifted by $\Delta\varphi/2=2\pi/4L$ relative to $\omega_{ij}$.
The poles therefore appear as pixels corners. In the next paragraphs,
we analyze the properties of equi-angular pixelizations which are
of interest in the implementation of scalar and spin $\pm2$ spherical
harmonics transforms. These properties are discussed in comparison
with the HEALPix pixelization%
\footnote{http://healpix.jpl.nasa.gov/%
} (Hierarchical Equal Area iso-Latitude Pixelization) \cite{SASgorski},
and the GLESP pixelization%
\footnote{http://www.glesp.nbi.dk/%
} (Gauss-Legendre Sky Pixelization) \cite{SASdoroshkevich1,SASdoroshkevich2},
which are widely used in astrophysics and cosmology.

Firstly, we discuss the asymptotic complexity for the computation
of scalar and spin $\pm2$ spherical harmonics transforms. Let us
consider band-limited functions $_{n}G(\omega)$ on the sphere with
band limit $L$, defined through the following condition on their
scalar ($n=0$) or spin-weighted ($n\neq0$) spherical harmonics coefficients:
$_{n}\widehat{G}_{lm}=0$ for $l\geq L$. For a signal with band-limit
$L$, the \emph{a priori} complexity associated with the naive computation
of the direct scalar spherical harmonics transform integral on the
sphere through simple discretization, \emph{i.e.} a quadrature, for
all $(l,m)$ with $\vert m\vert\leq l<L$, is naturally of order $\mathcal{O}(L^{4})$.
And the \emph{a priori} complexity associated with the naive computation
of the direct spin $\pm2$ spherical harmonics transforms integrals
on the sphere through simple quadrature, for all $(l,m)$ with $l\geq2$,
and $\vert m\vert\leq l<L$, is also naturally of order $\mathcal{O}(L^{4})$.
The same complexity naturally applies to the corresponding inverse
scalar or spin $\pm2$ transforms. We consider fine samplings corresponding
to megapixels maps on the sphere. In particular, the WMAP experiment
currently provides all-sky maps of around three megapixels. For such
a fine sampling defining a band limit around $L\simeq10^{3}$, the
typical computation time for $(2L)^{2}$ multiplications and $(2L)^{2}$
additions of double-precision numbers is of order of $0.03$ seconds
on a standard $2.2$ GHz Intel Pentium Xeon CPU. We take this value
as a fair estimation of the computation time required for one integration
for given $(l,m)$, or one summation for given $(\theta,\varphi)$,
with an associated $\mathcal{O}(L^{2})$ asymptotic complexity. Consequently,
scalar or spin $\pm2$ spherical harmonics transforms, with an asymptotic
complexity of order $\mathcal{O}(L^{4})$, typically take several
days at that band limit $L\simeq10^{3}$ on a single standard computer.
Considering the analysis of a large number of signals or simulations
may become difficultly affordable in terms of computation times, \emph{a
fortiori} in the perspective of forthcoming experiments with improved
resolution on the sky, such as the Planck satellite experiment which
will release all-sky maps of around fifty megapixels.

The development of a fast and exact algorithm is therefore of great
interest for the CMB analysis. The technique of separation of variables
in the scalar or spin $\pm2$ spherical harmonics into the associated
Legendre polynomials $P_{l}^{m}(\cos\theta)$ or the Wigner $d$-functions
$d_{m2}^{l}(\theta)$, and the complex exponentials $e^{im\varphi}$
allows to decompose the transform as successive transforms in $\varphi$
and $\theta$ \cite{SASmaslen1,SASmaslen2}. It naturally reduces
the asymptotic complexity for the direct and inverse scalar and spin
$\pm2$ spherical harmonics transforms to $\mathcal{O}(L^{3})$. It
can be implemented on any iso-latitude pixelization. Many pixelization
schemes have been considered on the sphere which satisfy this requirement.
It is the case for the equi-angular, HEALPix, and GLESP pixelizations.
The algorithms existing on HEALPix or GLESP pixelizations are indeed
based on this technique. As discussed in the next subsection, the
asymptotic complexity may be further reduced on equi-angular pixelizations.

Secondly, we discuss the precision of the computation. A sampling
theorem exists on equi-angular pixelizations on the sphere, which
represents a generalization of the Nyquist-Shannon theorem on the
line. The sampling theorem states that the scalar spherical harmonics
coefficients of a band-limited function on the sphere may be computed
exactly up to a band limit $L$, through a $2L\times2L$ equi-angular
sampling, as a finite weighted sum, \emph{i.e.} a quadrature, of the
sampled values of that function \cite{SASdriscoll}. The weights are
defined from the structure of the Legendre polynomials $P_{l}(\cos\theta)$
on the interval $[0,\pi]$. A Gaussian quadrature rule for the exact
computation of spherical harmonics coefficients of band-limited functions
also exists on GLESP pixelizations. The HEALPix implementation of
the scalar and spin $\pm2$ spherical harmonics transforms achieves
a very good precision thanks to an iteration process, but it is only
approximate from the theoretical point of view as no sampling theorem
is established on such pixelizations.

Thirdly, we comment on the notion of pixel window function. On equi-angular
pixelizations, the area $A$ of pixels varies considerably with the
co-latitude, from small pixels close to the poles, to larger pixels
around the equator: $A(\omega_{ij})\simeq\sin\theta_{i}\Delta\theta\Delta\varphi$.
This is a major difference with the HEALPix pixelization which defines
equal-area pixels, or the GLESP pixelization which defines nearly
equal-area pixels. The constant area of pixels is an important property
allowing the definition of a pixel window function associated to a
given pixelization at a given resolution. The main interest of this
concept is to apply a low-pass filtering to the signal, implementing
the fact that the pixelized signal is smoothed by integration over
the pixel area. The corresponding window function depends on the pixelization
structure and resolution. The procedure of pixelization is approximated
to a correlation of the signal with an axisymmetric beam, and therefore
strongly relies on the assumption of equal-area pixels. We do not
consider here the generalization of this concept on equi-angular pixelizations,
where the pixel area varies drastically over the surface of the sphere.
We only consider signals with band limit $L$ on a $2L\times2L$ equi-angular
sampling. In this case, for an application such as the downsampling,
the spherical harmonics coefficients of a signal can be computed exactly
thanks to the sampling theorem, and truncated at the desired band
limit. In that respect at least, the use of the pixel window function
can be avoided.

Let us finally emphasize that each pixelization scheme (equi-angular,
HEALPix, GLESP, ...) may provide specific advantages. The new algorithm
proposed in the next subsection on equi-angular pixelizations is exact
and has an asymptotic complexity of order $\mathcal{O}(L^{2}\log_{2}^{2}L)$.
But pixelizations with equal-area pixels represent an advantage when
dealing with noisy data \cite{SASgorski}. Our algorithm is therefore
to be understood as a simple alternative to the existing algorithms.
A more detailed comparison of the various algorithms is out of the
scope of the present work.

\subsection{New exact $\mathcal{O}(L^{2}\log_{2}^{2}L)$ algorithm \label{sub:New-exact-algo}}

We recall the following derivative relation on the associated Legendre
polynomials \cite{SASabramowitz},\begin{equation}
\left[\frac{\partial}{\partial\theta}P_{l}^{m}\right]\left(\cos\theta\right)=l\cot\theta P_{l}^{m}\left(\cos\theta\right)-\frac{l+m}{\sin\theta}P_{l-1}^{m}\left(\cos\theta\right),\label{12}\end{equation}
under the convention that $P_{l}^{m}$ is defined to be zero for $l<\vert m\vert$.
Through this relation, the derivative relations (\ref{10}) and (\ref{11})
between the spin $\pm2$ spherical harmonics $_{\pm2}Y_{lm}$ and
the scalar spherical harmonics $Y_{lm}$ may be turned into a simple
expression of $_{\pm2}Y_{lm}$ as linear combinations without derivatives
of $Y_{lm}$, $Y_{(l-1)m}$, and $Y_{(l-2)m}$. Notice that the same
recurrence procedure is used in a different context in \cite{SASwiaux2},
in order to express spin $n$ spherical harmonics $_{n}Y_{lm}$, for
any $n$ with $0\leq\vert n\vert\leq l$, as linear combinations of
scalar spherical harmonics. Through the recurrence relation on $l$
satisfied by the associated Legendre polynomials of given $m$,\begin{equation}
\left(l-m\right)P_{l}^{m}\left(\cos\theta\right)=\left(2l-1\right)\cos\theta P_{l-1}^{m}\left(\cos\theta\right)-\left(l+m-1\right)P_{l-2}^{m}\left(\cos\theta\right),\label{13}\end{equation}
the $Y_{(l-2)m}$ term in the quoted linear combination for $_{\pm2}Y_{lm}$
may be cancelled. We finally obtain the following expression of $_{\pm2}Y_{lm}$
as a linear combination of the scalar spherical harmonics $Y_{lm}$
and $Y_{(l-1)m}$:\begin{equation}
_{\pm2}Y_{lm}\left(\theta,\varphi\right)=\left[\frac{\left(l-2\right)!}{\left(l+2\right)!}\right]^{1/2}\left[\alpha_{(lm)}^{\pm}\left(\theta\right)Y_{lm}\left(\theta,\varphi\right)+\beta_{(lm)}^{\pm}\left(\theta\right)Y_{(l-1)m}\left(\theta,\varphi\right)\right],\label{14}\end{equation}
for $l\geq2$ and $\vert m\vert\leq l$, and with the functional coefficients\begin{eqnarray}
\alpha_{(lm)}^{\pm}\left(\theta\right) & = & \frac{2m^{2}-l\left(l+1\right)}{\sin^{2}\theta}\mp2m\left(l-1\right)\frac{\cot\theta}{\sin\theta}+l\left(l-1\right)\cot^{2}\theta\nonumber \\
\beta_{(lm)}^{\pm}\left(\theta\right) & = & 2\left[\frac{2l+1}{2l-1}\left(l^{2}-m^{2}\right)\right]^{1/2}\left(\pm\frac{m}{\sin^{2}\theta}+\frac{\cot\theta}{\sin\theta}\right).\label{15}\end{eqnarray}
This relation holds once more under the convention that $Y_{lm}$
is defined to be zero for $l<\vert m\vert$.

Consequently, the direct spin-weighted spherical harmonics transform
of a spin $\pm2$ function $_{\pm2}G$ may be written as a linear
combination of direct scalar spherical harmonics transforms for three
associated scalar functions. Indeed, if the associated functions are
defined by $G^{(p)}(\theta,\varphi)=(\cot^{p}\theta/\sin^{q}\theta){}_{\pm2}G(\theta,\varphi)$,
for $p,q\in\mathbb{N}$ and $p+q=2$, one gets from relation (\ref{14}):\begin{eqnarray}
_{\pm2}\widehat{G}_{lm} & = & \left[\frac{\left(l-2\right)!}{\left(l+2\right)!}\right]^{1/2}\left\{ 2\left[\frac{2l+1}{2l-1}\left(l^{2}-m^{2}\right)\right]^{1/2}\left(\widehat{G^{(1)}}_{(l-1)m}\pm m\widehat{G^{(0)}}_{(l-1)m}\right)\right.\nonumber \\
 &  & \left.+l\left(l-1\right)\widehat{G^{(2)}}_{lm}\mp2m\left(l-1\right)\widehat{G^{(1)}}_{lm}+\left[2m^{2}-l\left(l+1\right)\right]\widehat{G^{(0)}}_{lm}\right\} ,\nonumber \\
 & \ \label{16}\end{eqnarray}
with $l\geq2$ and $|m|\leq l$. The relation (\ref{14}) also implies
that the inverse spin-weighted transform of a set of spin $\pm2$
coefficients $_{\pm2}\widehat{G}_{lm}$ (with $_{\pm2}\widehat{G}_{lm}=0$
for $l\geq L$) may be written as a sum of three inverse scalar spherical
harmonics transforms:\begin{equation}
G\left(\theta,\varphi\right)=\frac{1}{\sin^{2}\theta}A\left(\theta,\varphi\right)+\frac{\cot\theta}{\sin\theta}B\left(\theta,\varphi\right)+\cot^{2}\theta C\left(\theta,\varphi\right),\label{17}\end{equation}
with the scalar functions $A$, $B$, and $C$ identified as follows
by their scalar spherical harmonics coefficients:\begin{eqnarray}
\widehat{A}_{lm} & = & \left[\frac{\left(l-2\right)!}{\left(l+2\right)!}\right]^{1/2}\left[2m^{2}-l\left(l+1\right)\right]{}_{\pm2}\widehat{G}_{lm}\nonumber \\
 &  & \pm2m\left[\frac{\left(l-1\right)!}{\left(l+3\right)!}\frac{\left(2l+3\right)}{\left(2l+1\right)}\left((l+1)^{2}-m^{2}\right)\right]^{1/2}{}_{\pm2}\widehat{G}_{(l+1)m}\nonumber \\
\widehat{B}_{lm} & = & \left[\frac{\left(l-2\right)!}{\left(l+2\right)!}\right]^{1/2}\left[\mp2m\left(l-1\right)\right]{}_{\pm2}\widehat{G}_{lm}\nonumber \\
 &  & +2\left[\frac{\left(l-1\right)!}{\left(l+3\right)!}\frac{\left(2l+3\right)}{\left(2l+1\right)}\left((l+1)^{2}-m^{2}\right)\right]^{1/2}{}_{\pm2}\widehat{G}_{(l+1)m}\nonumber \\
\widehat{C}_{lm} & = & \left[\frac{\left(l-2\right)!}{\left(l+2\right)!}\right]^{1/2}\left[l\left(l-1\right)\right]{}_{\pm2}\widehat{G}_{lm}.\label{18}\end{eqnarray}
As discussed, for functions band-limited at $L$, the separation of
variables allows to compute the direct and inverse scalar spherical
harmonics transforms in $\mathcal{O}(L^{3})$ operations. However,
a faster algorithm was developed by Driscoll and Healy on $2L\times2L$
equi-angular pixelizations on the sphere for the scalar spherical
harmonics transforms \cite{SASdriscoll}. The Fourier transforms in
$e^{im\varphi}$ are computed in $\mathcal{O}(L\log_{2}L)$ operations
for each $\theta$ through standard Cooley-Tukey fast Fourier transforms.
The algorithm also explicitly takes advantage of the recurrence relation
in $l$ on the associated Legendre polynomials $P_{l}^{m}(\cos\theta)$
to compute the direct associated Legendre transforms in $\mathcal{O}(L\log_{2}^{2}L)$
operations for each $m$. In these terms, the direct and inverse scalar
spherical harmonics transforms are computed in $\mathcal{O}(L^{2}\log_{2}^{2}L)$
operations. The computation is theoretically exact thanks to the sampling
theorem on equi-angular pixelizations. Corresponding stable numerical
implementations exist in the SpharmonicKit package \cite{SAShealy1,SAShealy2}%
\footnote{http://www.cs.dartmouth.edu/$\sim$geelong/sphere/%
}. Through the relations (\ref{16}) and (\ref{17}), the spin-weighted
spherical harmonics transform of a band-limited spin $\pm2$ function
with band limit $L$ may consequently also be computed exactly on
a $2L\times2L$ equi-angular pixelization on the sphere from the Driscoll
and Healy fast scalar spherical harmonics transforms, and with the
same asymptotic complexity of order $\mathcal{O}(L^{2}\log_{2}^{2}L)$.
In terms of our previous intuitive estimations, we recall that an
$\mathcal{O}(L^{2})$ scalar product requires the order of $0.03$
seconds on a standard $2.2$ GHz Intel Pentium Xeon CPU, at band limits
around $L\simeq10^{3}$. When compared to the \emph{a priori} $\mathcal{O}(L^{4})$
asymptotic complexity, the $\mathcal{O}(L^{2}\log_{2}^{2}L)$ scalar
and spin $\pm2$ spherical harmonics transforms algorithms consequently
reduce computation times from days to seconds for the fine samplings
considered.

Let us remark that a recurrence relation was proposed in \cite{SASkostelec1}
in order to compute spin $n$ spherical harmonics transforms from
scalar spherical harmonics transforms. However, the proposed relation
explicitly relates $_{n}Y_{lm}$ with $_{n\mp1}Y_{lm}$, $_{n\mp1}Y_{(l-1)m}$,
and $_{n\mp1}Y_{(l+1)m}$. The term $_{n\mp1}Y_{(l+1)m}$ increases
the band limit of the functions to be analyzed to $L+2$ after the
$2$-steps recurrence leading from spin $\pm2$ to scalar spherical
harmonics. On $2L\times2L$ equi-angular pixelizations, the SpharmonicKit
package is technically limited to consider coefficients lower than
$L$, and numerical errors will occur due to the absence of consideration
of the coefficients at $l=L$ and $l=L+1$. No such issue occurs from
the relation (\ref{14}) here above, which preserves the band limit
$L$ for the associated scalar functions.

\subsection{Numerical implementation}

We here report the computation times and memory requirements for the
numerical implementation of the algorithm at band limits up to $L=1024$,
and briefly discuss the issue of the numerical stability of the implementation.
The implementation is directly based on the fast scalar spherical
harmonics transform proposed by Driscoll and Healy and implemented
in the SpharmonicKit package. Computations are performed on a $2.20$
GHz Intel Pentium Xeon CPU with $2$ Gb of RAM memory. Random band-limited
test-functions are considered. Without loss of generality, these test-functions
are defined through their spin-weighted spherical harmonics coefficients
$_{\pm2}\widehat{G}_{lm}$, with $\vert m\vert\leq l<L$, and $l\geq2$,
with independent real and imaginary parts uniformly distributed in
the interval $[-1,+1]$. The inverse and direct spin-weighted spherical
harmonics transforms are successively computed, giving numerical coefficients
$_{n}\widehat{H}_{lm}$.

\begin{figure}
\begin{center}\includegraphics[width=6cm]{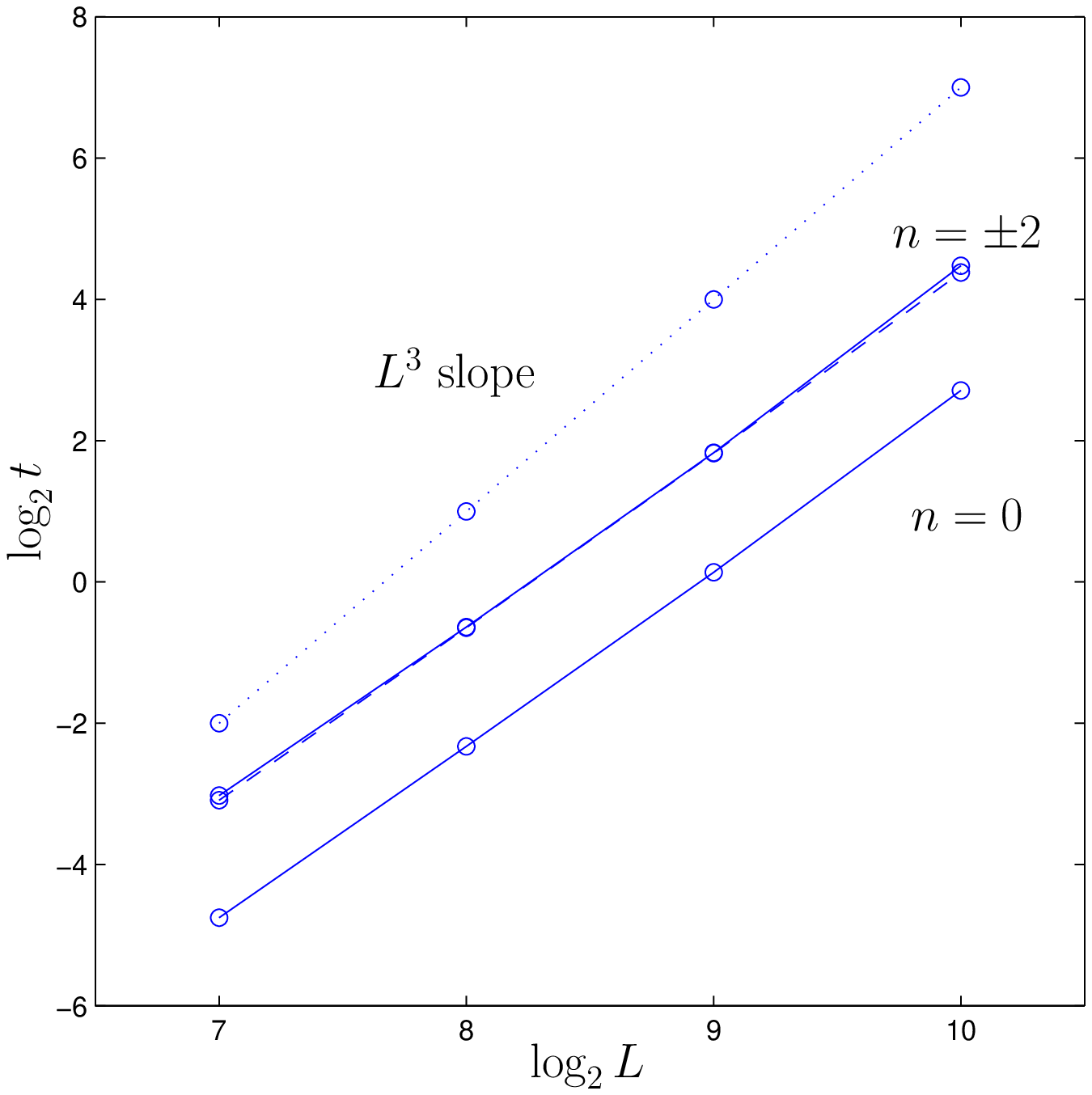}\hspace{1cm}\includegraphics[width=6cm]{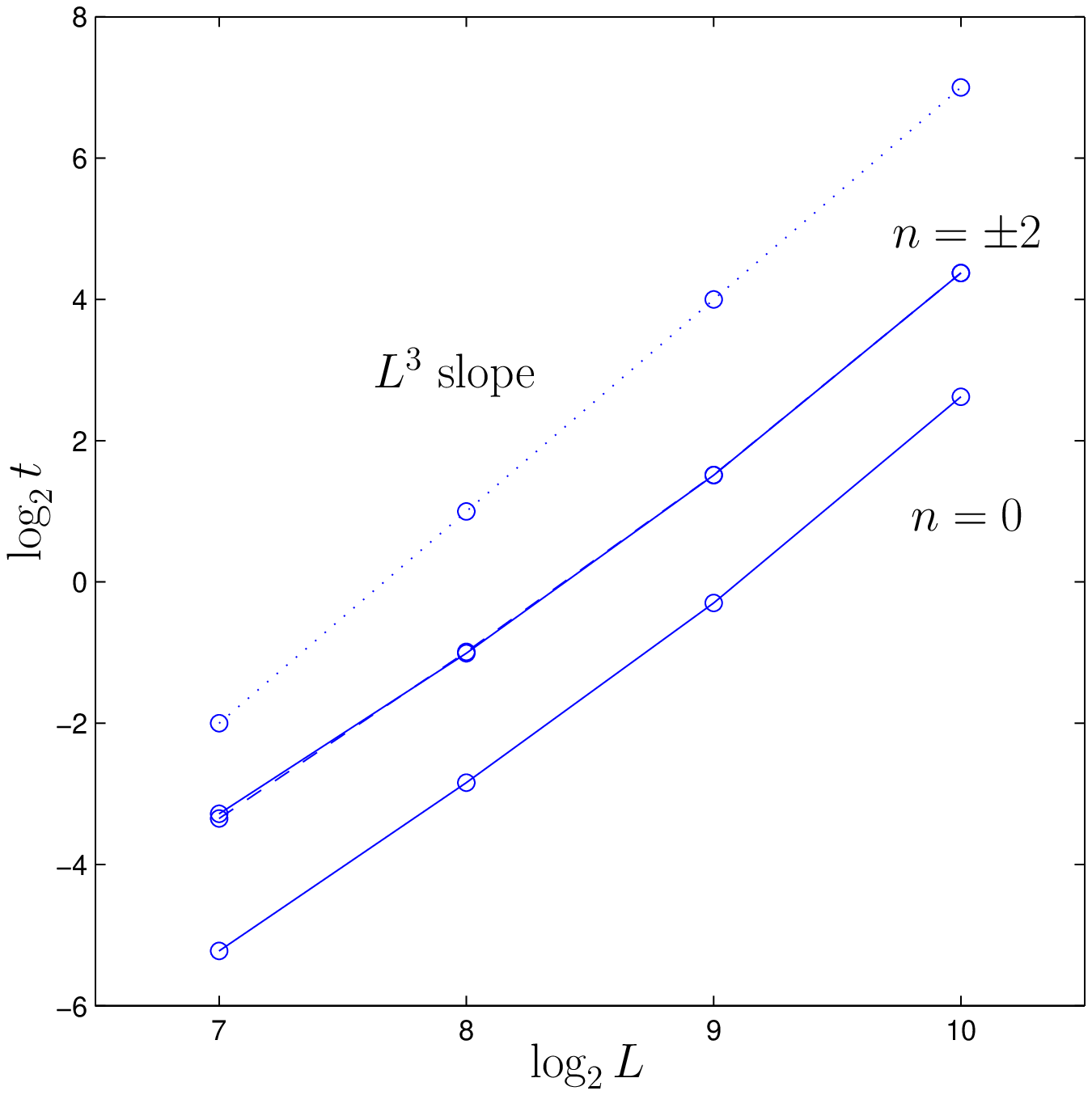}\end{center}

\caption{\label{cap:f2}Evolution of computation times $t$ displayed as $\log_{2}t-\log_{2}L$
for the direct (left) and inverse (right) spin-weighted spherical
harmonics transforms of spins $n=0$ (continuous line), $n=+2$ (continuous
line), and $n=-2$ (dashed line). Computation times are measured in
seconds on a $2.20$ GHz Intel Pentium Xeon CPU with $2$ Gb of RAM
memory, and reported for the band limits $L\in\{128,256,512,1024\}$.
The $\mathcal{O}(L^{2}\log^{2}L)$ asymptotic complexity is clearly
illustrated when compared to an $\mathcal{O}(L^{3})$ slope (dotted
line).}
\end{figure}

The computation times given in Table \ref{cap:t1} for the direct
and inverse spin $\pm2$ transforms are averages over $5$ random
test-functions. They range between $1.0\times10^{-1}$ seconds for
$L=128$ and $2.2\times10^{1}$ seconds for $L=1024$. The equality
of computation times for the positive and negative spins is an evident
consequence of the similarity of the $\pm2$ cases in the relation
(\ref{14}). The case $n=0$ corresponds to the scalar spherical harmonics
transform, and is added for comparison. The related values range between
$2.7\times10^{-2}$ seconds for $L=128$ and $6.5$ seconds for $L=1024$.
To summarize, computation times are of the order of seconds for a
band limit $L=1024$, in agreement with our previous intuitive estimations.
Both for the direct and inverse transforms, the evolution of the values
reported as a function of the band limit also supports the $\mathcal{O}(L^{2}\log_{2}^{2}L)$
behavior of the related asymptotic complexity, as illustrated in figure
\ref{cap:f2} in comparison with an $\mathcal{O}(L^{3})$ slope. The
ratio of computation times for the cases $n=\pm2$ and $n=0$ also
reflects the simple fact that three scalar transforms are computed
for each spin $\pm2$ transform.%
\begin{table}
\begin{center}\begin{tabular}{lcccc}
\hline 
Spin&
Time $L=128$&
Time $L=256$&
Time $L=512$&
Time $L=1024$\tabularnewline
&
(sec)&
(sec)&
(sec)&
(sec)\tabularnewline
\hline 
$n=0$&
$3.7\textrm{e}-02$&
$2.0\textrm{e}-01$&
$1.1\textrm{e}+00$&
$6.5\textrm{e}+00$\tabularnewline
&
$2.7\textrm{e}-02$&
$1.4\textrm{e}-01$&
$8.1\textrm{e}-01$&
$6.2\textrm{e}+00$\tabularnewline
\hline 
$n=2$&
$1.2\textrm{e}-01$&
$6.4\textrm{e}-01$&
$3.6\textrm{e}+00$&
$2.2\textrm{e}+01$\tabularnewline
&
$1.0\textrm{e}-01$&
$5.0\textrm{e}-01$&
$2.9\textrm{e}+00$&
$2.1\textrm{e}+01$\tabularnewline
\hline 
$n=-2$&
$1.2\textrm{e}-01$&
$6.4\textrm{e}-01$&
$3.5\textrm{e}+00$&
$2.1\textrm{e}+01$\tabularnewline
&
$1.0\textrm{e}-01$&
$5.0\textrm{e}-01$&
$2.9\textrm{e}+00$&
$2.1\textrm{e}+01$\tabularnewline
\hline
\end{tabular}\end{center}\vspace{0.5cm}

\caption{\label{cap:t1}Computation times for $n=0$ and $n=\pm2$ spherical
harmonics transforms measured on a $2.20$ GHz Intel Pentium Xeon
CPU with $2$ Gb of RAM memory. Times associated with the direct transforms
are listed above the corresponding times for the inverse transforms.}
\end{table}

In the present implementation based on the SpharmonicKit package,
the required associated Legendre polynomials $P_{l}^{m}(\cos\theta)$
are pre-calculated once for all values of $l$, $\theta$, and $m$,
and stored in RAM memory. The pre-computation time itself is of order
$\mathcal{O}(L^{3})$ through the use of a recurrence relation in
$l$ on the associated Legendre polynomials. This pre-computation
is by definition not taken into account in the reported computation
times, which consequently remain of order $\mathcal{O}(L^{2}\log_{2}^{2}L)$.
The number of real values of associated Legendre polynomials $P_{l}^{m}(\cos\theta)$
stored in RAM memory for all $l$, $\theta$, and $m$ is also of
order $\mathcal{O}(L^{3})$. The overall memory requirements allowing
the direct and inverse transforms with the present numerical implementation
correspondingly increase from $5.6$ Mb for $L=128,$ to $32$ Mb
for $L=256$, $220$ Mb for $L=512$, and $1.2$ Gb for $L=1024$.
These memory requirements are easily accessible on a single standard
computer.

The absolute and relative numerical errors are defined as $\max_{l,m}\vert\,\,{}_{n}\widehat{G}_{lm}-\,\,{}_{n}\widehat{H}_{lm}\vert$
and $\max_{l,m}\vert(\,\,{}_{n}\widehat{G}_{lm}-\,\,{}_{n}\widehat{H}_{lm})/\,\,{}_{n}\widehat{G}_{lm}\vert$
respectively, where $\vert\cdot\vert$ here denotes the complex norm,
and $n\in\{0,\pm2\}$. The numerical errors associated with the $\mathcal{O}(L^{2}\log_{2}^{2}L)$
spin-weighted spherical harmonics transforms given in Table \ref{cap:t2}
are averages for transforms over $5$ random band-limited test-functions.
Absolute and relative errors do not exceed the order of $8.4\times10^{-9}$
and $4.2\times10^{-7}$ respectively for band limits up to $L=1024$.
The $\mathcal{O}(L^{2}\log_{2}^{2}L)$ implementation of the spin
$\pm2$ spherical harmonics transforms is therefore stable for band
limits up to $L=1024$. The numerical stability of the algorithm might
also have been inferred from the corresponding stability of the Driscoll
and Healy fast direct scalar spherical harmonics transform algorithm,
tested for band limits up to $L=1024$ \cite{SAShealy1,SAShealy2}.
The only potential source of instability is related to the multiplication
factor $\cot^{p}\theta/\sin^{q}\theta$ defining the scalar functions
associated with a spin $\pm2$ function in the computation of a spin-weighted
spherical harmonics transform from the relation (\ref{14}). Each
such factor indeed corresponds to a division by $\sin^{2}\theta$,
which induces multiplications by numbers of the order of $L^{2}$
around the poles $\theta=\{0,\pi\}$, where $L$ is the band limit
considered. However such operations could only produce numerical instabilities
for very high band limits, and obviously remain completely safe at
$L=1024$.%
\begin{table}
\begin{center}\begin{tabular}{lcccc}
\hline 
Spin&
Error $L=128$&
Error $L=256$&
Error $L=512$&
Error $L=1024$\tabularnewline
\hline 
$n=0$&
$1.8\textrm{e}-10$&
$6.5\textrm{e}-10$&
$2.3\textrm{e}-09$&
$8.4\textrm{e}-09$\tabularnewline
&
$9.7\textrm{e}-10$&
$5.7\textrm{e}-09$&
$1.6\textrm{e}-08$&
$1.1\textrm{e}-07$\tabularnewline
\hline 
$n=2$&
$1.8\textrm{e}-10$&
$6.6\textrm{e}-10$&
$2.4\textrm{e}-09$&
$8.3\textrm{e}-09$\tabularnewline
&
$7.2\textrm{e}-10$&
$4.2\textrm{e}-09$&
$4.6\textrm{e}-08$&
$4.2\textrm{e}-07$\tabularnewline
\hline 
$n=-2$&
$1.8\textrm{e}-10$&
$6.6\textrm{e}-10$&
$2.3\textrm{e}-09$&
$8.3\textrm{e}-09$\tabularnewline
&
$9.8\textrm{e}-10$&
$2.9\textrm{e}-09$&
$3.1\textrm{e}-08$&
$1.2\textrm{e}-07$\tabularnewline
\hline
\end{tabular}\end{center}\vspace{0.5cm}

\caption{\label{cap:t2}Errors are measured on a $2.20$ GHz Intel Pentium
Xeon CPU with $2$ Gb of RAM memory. Absolute errors after inverse
and direct transforms are listed above the corresponding relative
errors.}
\end{table}

\section{Application in cosmology}

\label{sec:CMB-polarization-spectra}In this section, we illustrate
the interest of the algorithm presented in the previous section in
the context of the analysis of CMB polarization data. The discussion
is based on the following introductory papers \cite{CMBPOLkosowsky1,CMBPOLzaldarriaga,CMBPOLseljak,CMBPOLkamionkowski1,CMBPOLkamionkowski2,CMBPOLhu1}
and reviews \cite{CMBPOLhu2,CMBPOLkosowsky2,CMBPOLcabella,CMBPOLlin}
relative to the CMB polarization analysis.

\subsection{Stokes parameters}

The CMB is observed in each direction $\omega=(\theta,\varphi)$ of
the sky as an incoming radial radiation, to which is associated a
transverse electromagnetic field thus lying in the tangent plane to
the sphere at the point considered. In that plane, we consider the
basis $(\hat{e}_{\theta},\hat{e}_{\varphi})$ with $\hat{e}_{\theta}$
pointing in the direction of increasing $\theta$ along each meridian,
and $\hat{e}_{\varphi}$ in the direction of increasing $\varphi$
along each parallel. In this so-called linear polarization basis,
nearly monochromatic radiation around a frequency $\omega_{r}$ may
be decomposed as an electric field with components $E_{\theta}(\omega_{r},t)=\mathcal{R}e[\varepsilon_{\theta}(t)e^{-i\omega_{r}t}]$
and $E_{\varphi}(\omega_{r},t)=\mathcal{R}e[\varepsilon_{\varphi}(t)e^{-i\omega_{r}t}]$.
The complex amplitudes $\varepsilon_{\theta}(t)$ and $\varepsilon_{\varphi}(t)$
slowly vary in time relatively to the timescale set by the wave period.
The intensity matrix $\mathbf{I}$ associated with the radiation simply
reads as the time average of the electric field rank $2$ tensor $[\varepsilon_{i}^{*}(t)\varepsilon_{j}(t)]\hat{e}_{i}\otimes\hat{e}_{j}$,
for $i,j\in\{\theta,\varphi\}$ \cite{CMBPOLkosowsky1,CMBPOLhu1}.
It thus naturally decomposes on the $2\times2$ matrix basis formed
by the identity matrix $\sigma_{0}=\mathbb{I}$, and the well-known
Pauli matrices $(\sigma_{1},\sigma_{2},\sigma_{3})$, as $\mathbf{I}=[I\sigma_{0}+U\sigma_{1}+V\sigma_{2}+Q\sigma_{3}]/2$.
The constants $I$, $U$, $V$, and $Q$ define the four real Stokes
parameters \cite{CMBPOLjackson} associated with the radiation: $I=\langle\vert\varepsilon_{\theta}(t)\vert^{2}+\vert\varepsilon_{\varphi}(t)\vert^{2}\rangle$
, $Q=\langle\vert\varepsilon_{\theta}(t)\vert^{2}-\vert\varepsilon_{\varphi}(t)\vert^{2}\rangle$,
$U=\langle\varepsilon_{\theta}^{*}(t)\varepsilon_{\varphi}(t)+\varepsilon_{\theta}(t)\varepsilon_{\varphi}^{*}(t)\rangle$,
$V=i\langle\varepsilon_{\theta}^{*}(t)\varepsilon_{\varphi}(t)-\varepsilon_{\theta}(t)\varepsilon_{\varphi}^{*}(t)\rangle$.
The brackets $\langle\cdot\rangle$ denote time averaging. If the
two components $\varepsilon_{\theta}(t)$ and $\varepsilon_{\varphi}(t)$
are correlated, the radiation is said to be polarized. The positive
parameter $I$ may be identified with the overall intensity of radiation,
while $Q$ and $U$ identify with the linear polarizations, and $V$
with the circular polarization. Unpolarized radiation, or natural
light, is therefore characterized by $Q=U=V=0.$

As functions on the sphere, $I(\omega)$, $Q(\omega)$, $U(\omega)$
and $V(\omega)$ have different behaviors both under parity, \emph{i.e.}
global inversion ($\cdot''$) of the coordinates, and under local
rotations ($\cdot'$) of the basis vectors $(\hat{e}_{\theta},\hat{e}_{\varphi})$
in the tangent plane at $\omega=(\theta,\varphi)$. A global inversion
of the right-handed three-dimensional Cartesian coordinate system
$(o,o\hat{x},o\hat{y},o\hat{z})$ centered on the unit sphere induces
the following modification of Cartesian coordinates: $(x'',y'',z'')=(-x,-y,-z)$.
The spherical coordinates $\omega=(\theta,\varphi)$ of a given point
on $S^{2}$ change according to $\omega''=(\theta'',\varphi'')=(\pi-\theta,\pi+\varphi)$.
Locally in the tangent plane, the global inversion also implies an
inversion of the basis vector $\hat{e}_{\theta}$: $(\hat{e}_{\theta}'',\hat{e}_{\varphi}'')=(-\hat{e}_{\theta},\hat{e}_{\varphi})$.
The Stokes parameters $I$ and $Q$ have even parity, $I''(\omega'')=I(\omega)$
and $Q''(\omega'')=Q(\omega)$, while $U$ and $V$ have odd parity,
$U''(\omega'')=-U(\omega)$ and $V''(\omega'')=-V(\omega)$. Under
local rotations of the basis vectors $(\hat{e}_{\theta},\hat{e}_{\varphi})$
by an angle $\chi_{0}$, the coordinates $\vec{\varepsilon}=(\varepsilon_{\theta},\varepsilon_{\varphi})$
of vectors in the tangent plane transform through $\vec{\varepsilon}\,'=r_{\chi_{0}}\cdot\vec{\varepsilon}$,
for the standard rotation matrix $r_{\chi_{0}}$, with entries $r_{\chi_{0}}^{11}=r_{\chi_{0}}^{22}=\cos\chi_{0}$
and $r_{\chi_{0}}^{12}=-r_{\chi_{0}}^{21}=\sin\chi_{0}$. The Stokes
parameters $I$ and $V$ are invariant while $Q$ and $U$ are mixed
by local rotations. Equivalently, one may also rewrite the intensity
matrix as\begin{equation}
\mathbf{I}=\frac{1}{2}\left[I\sigma_{0}+V\sigma_{2}+\left(Q+iU\right)\sigma_{+}+\left(Q-iU\right)\sigma_{-}\right],\label{19}\end{equation}
with $\sigma_{\pm}=(\sigma_{3}\mp i\sigma_{1})/2$. The Pauli matrices
transform as $\sigma'_{\mu}=r_{\chi_{0}}\cdot\sigma_{\mu}\cdot r_{\chi_{0}}^{T}$,
for $\mu=\{0,1,2,3\}$. The matrices $\sigma_{0}$ and $\sigma_{2}$
are thus invariant, while $\sigma_{\pm}$ transform as $\sigma_{\pm}'=e^{\mp2i\chi_{0}}\sigma_{\pm}$.
Consequently the four Stokes parameters are associated with spin functions
on the sphere. The intensity $I(\omega)$ and the circular polarization
parameter $V(\omega)$ are scalar functions. The combinations $(Q\pm iU)(\omega)$
are spin $\pm2$ functions: $(Q\pm iU)'(\omega)=e^{\mp2i\chi_{0}}(Q\pm iU)(\omega)$.
Notice that under parity these two combinations transform in one another:
$(Q\pm iU)''(\omega'')=(Q\mp iU)(\omega)$ \cite{CMBPOLzaldarriaga,CMBPOLhu1}.

\subsection{Angular power spectra}

It is assumed that the physics of the CMB is invariant under parity
and under local rotations. It is therefore suitable to relate the
observables $I$, $Q$, $U$, and $V$ to invariant physical quantities.
The intensity $I(\omega)$ defines the CMB temperature anisotropies
$T(\omega)$ and is indeed itself invariant under the transformations
considered. As no circular polarization may arise from Thomson scattering,
the CMB polarization is completely described in terms of the two linear
polarization Stokes parameters $Q$ and $U$. It is equivalently defined
by their spin $\pm2$ combinations $Q\pm iU$. Associated polarization
components, real scalar functions on the sphere and parity eigenmodes,
are naturally defined from $Q\pm iU$ in terms of the raising $\eth$
and lowering $\bar{\eth}$ operators respectively given in (\ref{5})
and (\ref{6}). These components $\tilde{E}(\omega)=-[\bar{\eth}^{2}(Q+iU)(\omega)+\eth^{2}(Q-iU)(\omega)]/2$,
and $\tilde{B}(\omega)=i[\bar{\eth}^{2}(Q+iU)(\omega)-\eth^{2}(Q-iU)(\omega)]/2$,
have even and odd parities and are therefore referred to as electric
and magnetic components respectively \cite{CMBPOLzaldarriaga}. Let
us consider the decomposition of the spin $\pm2$ functions $Q\pm iU$
in spin-weighted spherical harmonics and the relations $_{2}Y_{lm}=N_{(l2)}\eth^{2}\, Y_{lm}$
and $_{-2}Y_{lm}=N_{(l2)}\bar{\eth}^{2}\, Y_{lm}$, with $N_{(l2)}=[(l-2)!/(l+2)!]^{1/2}$,
induced from (\ref{10}) and (\ref{11}). The application of the raising
and lowering operators on this decomposition through the relations
(\ref{7}) to (\ref{9}) gives $\widehat{\tilde{E}}_{lm}=\widehat{E}_{lm}/N_{(l2)}$
and $\widehat{\tilde{B}}_{lm}=\widehat{B}_{lm}/N_{(l2)}$, where\begin{equation}
\widehat{E}_{lm}=-\frac{1}{2}\left(\,\,{}_{+2}\widehat{\left(Q+iU\right)}_{lm}+\,\,{}_{-2}\widehat{\left(Q-iU\right)}_{lm}\right)\label{20}\end{equation}
 and \begin{equation}
\widehat{B}_{lm}=\frac{i}{2}\left(\,\,{}_{+2}\widehat{\left(Q+iU\right)}_{lm}-\,\,{}_{-2}\widehat{\left(Q-iU\right)}_{lm}\right),\label{21}\end{equation}
define the properly normalized real $E(\omega)$ and $B(\omega)$
components. These coefficients are explicitly invariant under local
rotations.

The random process from which the CMB radiation arises is assumed
to be Gaussian and stationary. It is therefore completely characterized
in terms of its temperature and polarization two-point correlation
functions. The corresponding invariant angular power spectra are naturally
those associated with the temperature $TT$, the polarizations $EE$
and $BB$, and the cross-correlation between the temperature and electric
polarization component $TE$: $\langle\widehat{T}_{l'm'}^{*}\widehat{T}_{lm}\rangle=C_{l}^{TT}\delta_{ll'}\delta_{mm'}$,
$\langle\widehat{E}_{l'm'}^{*}\widehat{E}_{lm}\rangle=C_{l}^{EE}\delta_{ll'}\delta_{mm'}$,
$\langle\widehat{B}_{l'm'}^{*}\widehat{B}_{lm}\rangle=C_{l}^{BB}\delta_{ll'}\delta_{mm'}$,
$\langle\widehat{T}_{l'm'}^{*}\widehat{E}_{lm}\rangle=C_{l}^{TE}\delta_{ll'}\delta_{mm'}$.
These physical quantities are indeed invariant under local rotations
and parity. The $TB$ and $EB$ cross-correlations are specifically
excluded from the requirement of invariance under parity.

Notice that the $E$ and $B$ components of polarization not only
define invariant physical angular power spectra, but they are also
associated with different mechanisms of production of the radiation,
corresponding to different theoretical cosmological models. Scalar
primordial energy density perturbations only produce the $E$ polarization
component, while vector and tensor (\emph{i.e.} gravity waves) perturbations
produce both $E$ and $B$ polarization components.

\subsection{Numerical illustration}

Scalar and spin $\pm2$ direct spherical harmonics transforms are
required for the estimation of the CMB angular power spectra from
the observables $T$, $Q$, and $U$ \cite{CMBPOLzaldarriaga}. The
simulation of temperature and polarization maps from given theoretical
angular power spectra requires the corresponding inverse transforms.
We apply our algorithm to simulate CMB maps and angular power spectra,
for illustration of its precision and speed performances.

\begin{figure}
\begin{center}\includegraphics[width=6cm]{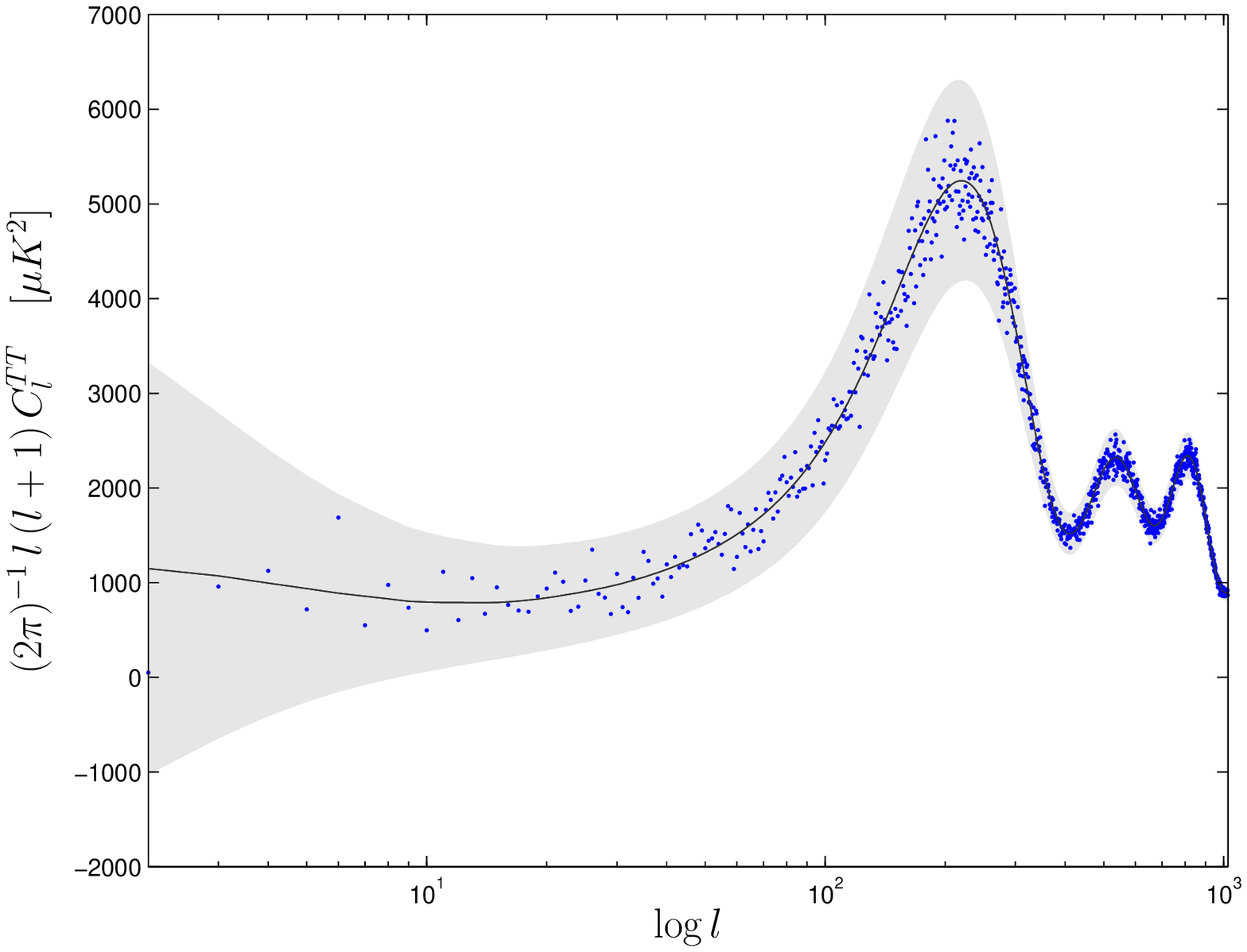}\\
\includegraphics[width=6cm]{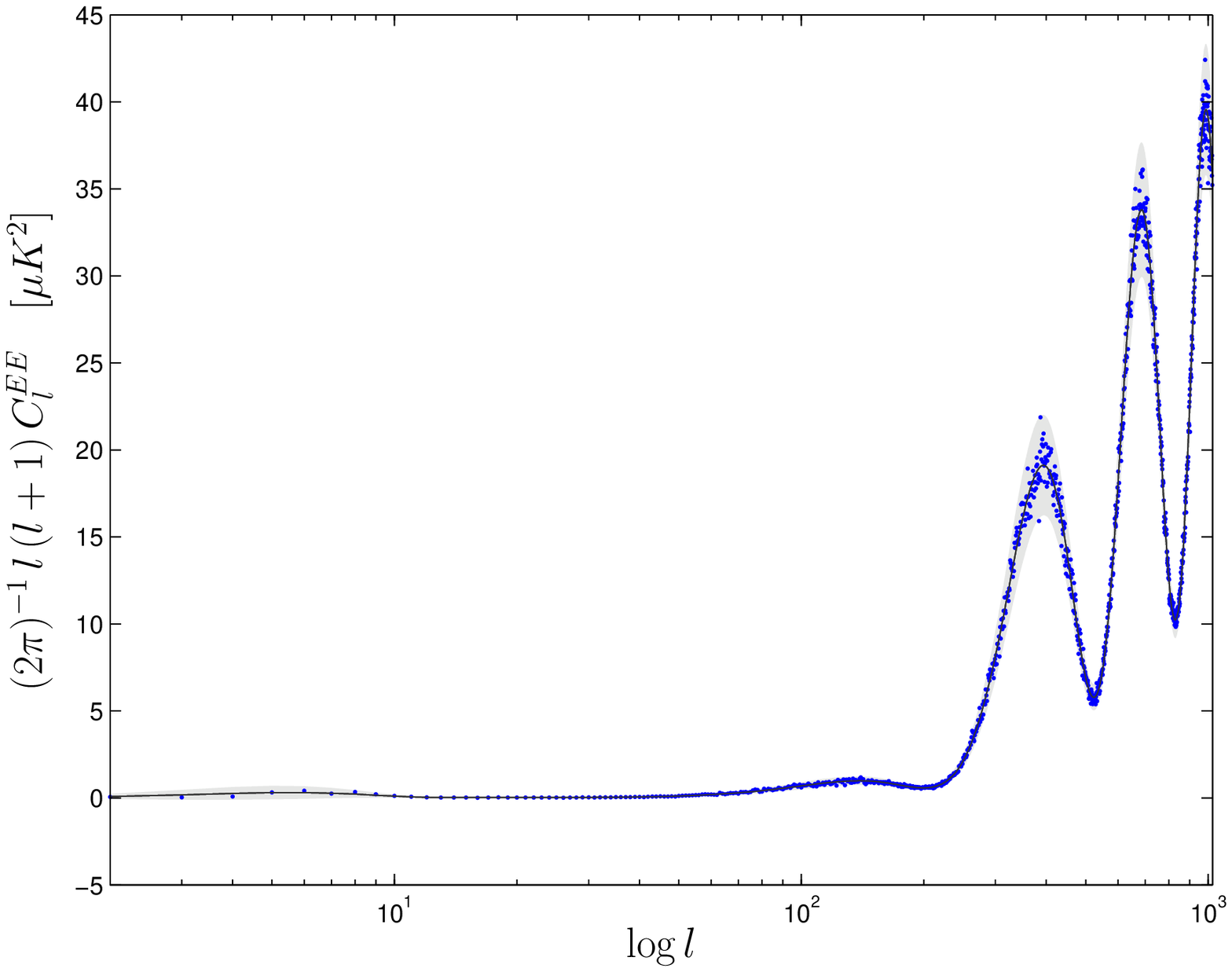}\hspace{1cm} \includegraphics[width=6cm]{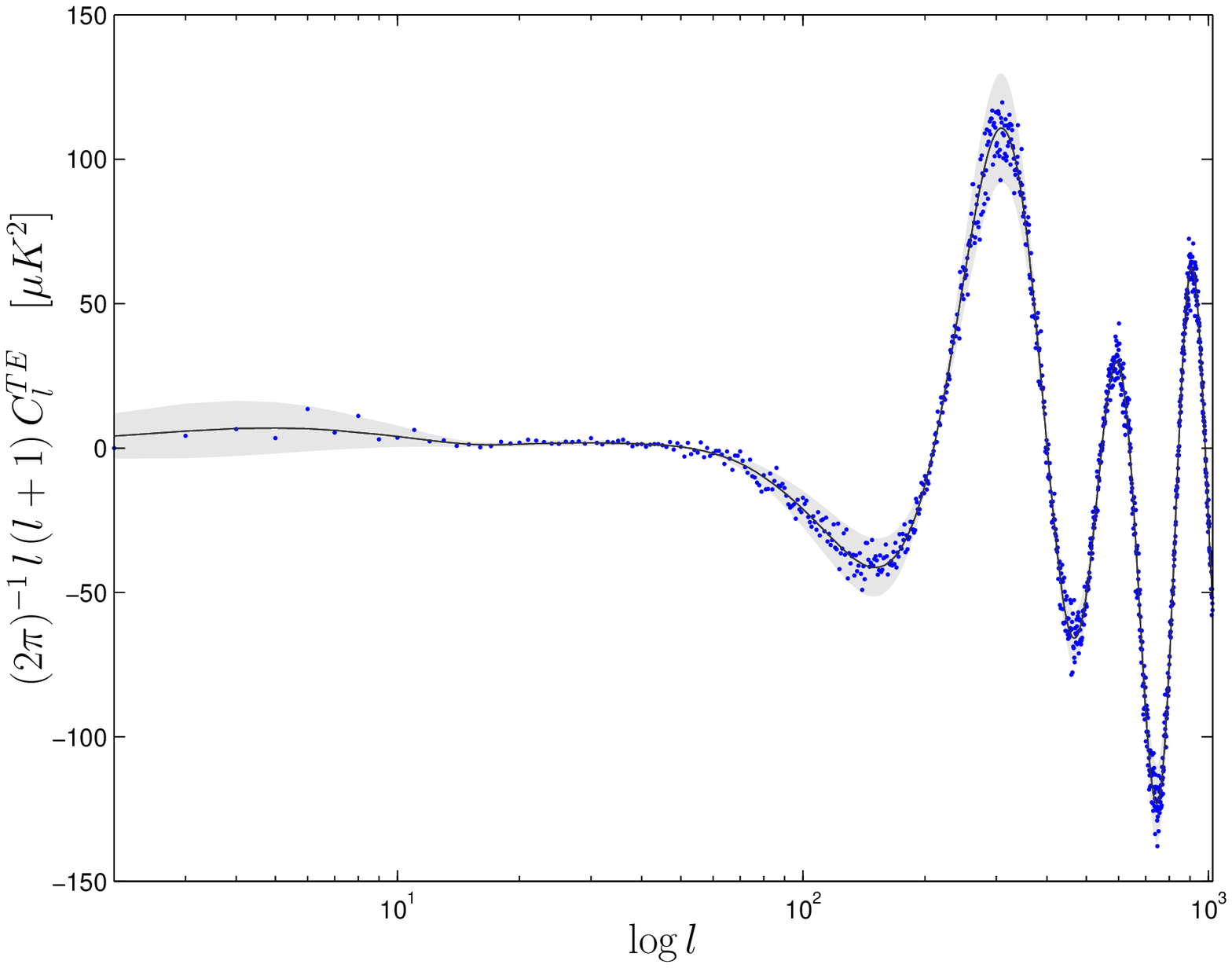}\end{center}

\caption{\label{cap:f4}CMB temperature and polarization angular power spectra
$C_{l}^{TT}$ (top), $C_{l}^{EE}$(bottom left), and $C_{l}^{TE}$
(bottom right) of the cosmic microwave background up to a band limit
$L=1024$ and in $\mu K^{2}$. Inverse and direct transforms are successively
performed through the exact $\mathcal{O}(L^{2}\log_{2}^{2}L)$ scalar
and spin $\pm2$ spherical harmonics transforms on $2L\times2L$ equi-angular
pixelizations on the sphere, in order to produce the estimated spectra
(scattered points) from the original spectra of the concordance cosmological
model (continuous lines). The original and estimated spectra coincide
within the $3\sigma$ uncertainty defined by the cosmic variance (grey
region).}
\end{figure}

We start from the temperature and polarization spectra $C_{l}^{TT}$,
$C_{l}^{EE}$, and $C_{l}^{TE}$ defined by the concordance cosmological
model which best fits the three-year WMAP data (the $BB$ polarization
spectrum is identically null: $C_{l}^{BB}=0$ ). These spectra are
represented in figure \ref{cap:f4} up to a band limit $L=1024$.
Spherical harmonics coefficients $\widehat{T}_{lm}$ and $\widehat{E}_{lm}$
are built up as the two marginal complex Gaussian realizations arising
from a jointly Gaussian statistical distribution with variances $C_{l}^{TT}$
and $C_{l}^{EE}$, and a covariance $C_{l}^{TE}$. The $T$, $Q$,
and $U$ maps are then produced by inverse scalar and spin $\pm2$
transforms, through the relations (\ref{20}) and (\ref{21}), with
$\widehat{B}_{lm}=0$.%

From the maps obtained, we recompute spherical harmonics coefficients
$\widehat{T}'_{lm}$, $\widehat{E}'_{lm}$, and $\widehat{B}'_{lm}$
by direct scalar and spin $\pm2$ transforms. Within the numerical
accuracy of the computer, the $B$ polarization coefficients are identically
null, in perfect agreement with the original data: $\widehat{B}'_{lm}=0$.
We finally estimate the temperature and polarization angular power
spectra from those coefficients as: $C_{l}^{TT}{}'=\sum_{m=-l}^{l}\vert\widehat{T}'_{lm}\vert^{2}/(2l+1)$,
$C_{l}^{EE}{}'=\sum_{m=-l}^{l}\vert\widehat{E}'_{lm}\vert^{2}/(2l+1)$,
$C_{l}^{TE}{}'=\sum_{m=-l}^{l}\widehat{T}'{}_{lm}^{*}\widehat{E}'_{lm}/(2l+1)$,
and $C_{l}^{BB}{}'=\sum_{m=-l}^{l}\vert\widehat{B}'_{lm}\vert^{2}/(2l+1)=0$.
These estimators follow chi-square distributions with $2l+1$ degrees
of freedom. For $X\in\{ TT,EE,BB,TE\}$, this induces a fractional
uncertainty $\sigma_{C_{l}^{X}{}'}/C_{l}^{X}=[2/(2l+1)]^{1/2}$ in
the estimation. This cosmic variance is large at low $l$ and small
at high $l$. The figure \ref{cap:f4} represents the good coincidence
between the original and estimated spectra up to the corresponding
uncertainty at each $l$. The computation time associated with the
overall procedure is $150$ seconds on a $2.20$ GHz Intel Pentium
Xeon CPU with $2$ Gb of RAM memory. In summary, this application
illustrates the good precision and speed performances of our fast
and exact algorithm, coherently with the results of tables \ref{cap:t1}
and \ref{cap:t2}.

\section{Conclusion}

\label{sec:Conclusion}In conclusion, we developed a fast, exact,
and stable algorithm for the spin $\pm2$ spherical harmonics transforms
of band-limited functions with band limit $L$ on $2L\times2L$ equi-angular
pixelizations on the sphere. The algorithm is based the Driscoll and
Healy fast scalar spherical harmonics transform algorithm. The exactness
of the computation on equi-angular pixelizations relies on a sampling
theorem. The associated asymptotic complexity is of order $\mathcal{O}(L^{2}\log_{2}^{2}L)$.
The algorithm is presented as an alternative to existing algorithms
with an asymptotic complexity of order $\mathcal{O}(L^{3})$ on the
HEALPix and GLESP pixelizations, which are widely used in the context
of astrophysics and cosmology.

The numerical implementation produced confirms the characteristics
of the algorithm. Typical computation times for $L=1024$ are of the
order of seconds. We also illustrated the interest of the algorithm
in the context of the analysis of CMB polarization data.

\begin{ack}
The authors wish to thank P. Vielva, J.-P. Antoine, B. Barreiro, and
E. Mart\'inez-Gonz\'alez for valuable comments and discussions.
Y. W. acknowledges support of the Swiss National Science Foundation
(SNF) under contract No. 200021-107478/1. He is also postdoctoral
researcher of the Belgian National Science Foundation (FRS-FNRS).
L. J. is postdoctoral researcher of the Belgian National Science Foundation
(FRS-FNRS).
\end{ack}

\end{document}